%


\input phyzzx
\hfuzz 25pt

\def\dplus{=\hskip-5pt \raise 0.7pt\hbox{${}_\vert$} ^{\phantom 7}}
\def\dplusup{=\hskip-5.1pt \raise 5.4pt\hbox{${}_\vert$} ^{\phantom 7}}
\def\dplus{=\hskip-4.8pt \raise 0.7pt\hbox{${}_\vert$} ^{\phantom 7}}
\def\dpar{\partial\hskip-6pt \raise 0.7pt\hbox{$/$} ^{\phantom 7}}

\def\pmb#1{\setbox0=\hbox{#1} \kern-.025em\copy0\kern-\wd0
\kern0.05em\copy0\kern-\wd0 \kern-.025em\raise.0433em\box0}

\def\i{\iota}

\font\mybb=msbm10 at 10pt

\def\bb#1{\hbox{\mybb#1}}

\def\a{\alpha}
\def\b{\beta}
\font\mybb=msbm10 at 12pt
\def\bb#1{\hbox{\mybb#1}}
\def\Z {\bb{Z}}
\def\R {\bb{R}}
 \def\unit{\hbox to 3.3pt{\hskip1.3pt \vrule height 7pt width .4pt \hskip.7pt
\vrule height 7.85pt width .4pt \kern-2.4pt
\hrulefill \kern-3pt
\raise 4pt\hbox{\char'40}}}

\REF\strom{A. Strominger, {\sl Open p-branes}, Phys. Lett. 
{\bf B383} 44; hep-th/9512059.}
\REF\town{P.K. Townsend, {\sl Brane Surgery}, Nucl. 
Phys. Proc. Suppl. {\bf 58} (1997) 163.}
\REF\pt{G. Papadopoulos and P.K. Townsend, 
{\sl Intersecting M-branes}, Phys. Lett. {\bf B380}
(1996) 273.}
\REF\callan {C.G. Callan, Jr and J.M. Maldacena, 
{\sl Brane dynamics from the
Born-Infeld action}, hep-th/9708147. }
\REF\howeb{P.S. Howe, N.D. Lambert and P. West, 
{\sl The self-dual string soliton},
hep-th/9709014.} 
\REF\howe{P.S. Howe, N.D. Lambert and P. West, 
{\sl The  three-brane soliton of the M-five-brane},
hep-th/9710033; {\sl Classical M-five-brane dynamics 
and quantum N=2 Yang-Mills}, hep-th/9710034.}
\REF\gibbons{G.W. Gibbons, {\sl Born-Infeld Particles 
and Dirichlet p-branes}, hep-th/9709027.}
\REF\pol{E. Witten and J. Polchinski, {\sl Evidence for Heterotic Type I 
String Duality}, Nucl. Phys. {\bf B460} (1996) 525.}
\REF\eric{ E. Bergshoeff, M. B. Green, G. Papadopoulos, M. de Roo and P.K. 
Townsend, {\sl Duality of Type II 7-branes and 8-branes}, Nucl. Phys. 
{\bf B470} (1996) 113.}
\REF\extra{ P.K. Townsend, {\sl World Sheet Electromagnetism and the 
Superstring Tension},
Phys. Lett. {\bf 277B} (1992) 285.}
\REF\ericpaul{E. Bergshoeff, L.A.J. London and P.K. Townsend, {\sl Spacetime 
Scale Invariance and
the Super-p-brane}, Class. Quantum Grav. {\bf 9} (1992) 25, hep-th/9206026. }
\REF\ericpaulb{E.B. Bergshoeff, J. Gomis and P.K. Townsend, 
{\sl M-brane Intersections from
Worldvolume Superalgebras}, hep-th/9711043, to appear in Phys. Lett. B.}
\REF\papac{E. Bergshoeff, R. Kallosh, 
T. Ort\'in and G. Papadopoulos, {\sl $\kappa$-Symmetry,
Supersymmetry and Intersecting Branes}, Nucl. Phys. {\bf B502} (1997) 149,
hep-th/9705040.}
\REF\papanew{G. Papadopoulos, {\sl T-duality and the Worldvolume Solitons of 
Five-Branes
and KK-Monopoles}, hep-th/9712162.}
\REF\hull{C.M. Hull, {\sl Gravitational Duality, Branes and Charges}, 
Nucl. Phys. {\bf B509} (1998) 216, hep-th/9705162.}
\REF\paulmartin{P.K. Townsend, {\sl Membrane Tension 
and Manifest IIB S-duality}, {\tt hep-th/9705160}; M. Cederwall and
P.K. Townsend, {\sl The Manifestly $SL(2,\Z)$-Covariant Superstring}, 
JHEP {\bf 09} (1997) 3.}
\REF\ericpaulc{E. Bergshoeff and P.K. Townsend, 
{\sl Super D-branes}, Nucl. Phys. {\bf B490} (1997)
145.}
\REF\nil{M. Cederwall, A. von Gussich, B.E.W. 
Nilsson, P. Sundell and A. Westerberg,
{\sl The Dirichlet Super P-Branes in Ten-Dimensional
 Type IIA and IIB Supergravity}, Nucl. Phys.
{\bf B490} (1997) 179.}
\REF\aga{M. Aganagic, C. Popescu and J.H. Schwarz, 
{\sl D-Brane Actions with Local Kappa
Symmetry}, Phys. Lett. {\bf B393} 311.}
\REF\gates{S.J. Gates, Jr., {\sl Super p-Form gauge Superfields}, 
Nucl. Phys. {\bf B184} (1981)
381.}
 
\Pubnum{ \vbox{ \hbox{R/98/03}\hbox{UG-1/98} } }
\pubtype{}
\date{January, 1998}
\titlepage
\title{Domain Walls on the Brane}
\author{E. Bergshoeff,  J.P. van der Schaar}
\address{\negthinspace Institute for Theoretical Physics \break 
University of Groningen\break
Nijenborgh 4, 9747 AG Groningen, The Netherlands}
\andauthor{ G. Papadopoulos}
\address{\negthinspace DAMTP \break Silver Street \break Cambridge CB3 9EW}
\abstract {
We show that all branes admit worldvolume domain
wall solutions.  We find one class of
solutions for which the tension of the brane changes 
discontinuously along the domain wall. These
solutions are not supersymmetric. We 
argue that there is another class of  domain
wall solutions which is supersymmetric. A particular
 case concerns supersymmetric domain wall
solutions on IIB D-5- and NS-5-branes.  }

\endpage
\pagenumber=2
\font\mybb=msbm10 at 12pt
\def\bb#1{\hbox{\mybb#1}}

\def\Z{\bb{Z}}
\def\R{\rlap I\mkern3mu{\rm R}}

\def\C{\mkern1mu\raise2.2pt\hbox{$\scriptscriptstyle|$}\mkern-7mu{\rm C}}

\def\i{\iota}
\def\l {\lambda}
\def\a{\alpha}
\def\b{\beta}
\def\k{\kappa}
\def\p {\phi}

\sequentialequations

\chapter{Introduction}

The investigation of the worldvolume solitons of 
various branes have given new insights
in the understanding of the properties of M-theory 
and string theory. From the bulk perspective, the worldvolume
solitons on a brane have the interpretation as the intersection region of two
 branes or the boundary of a brane ending on another [\strom-\pt]. As such
the worldvolume solitons themselves have a brane interpretation.
Many such worldvolume solitons have already been found [\callan-\gibbons] and 
perhaps the most notable one
is the 0-brane soliton on a D-brane which can be thought as the boundary
of a fundamental string ending on it [\gibbons]. 

A special class of solitons are domain walls. 
One example of a domain wall is the D-8-brane of massive IIA supergravity which
separates spacetime in two disconnected regions [\pol, \eric]. The
 D-8-brane is associated
with a ten-form field strength whose dual is the spacetime cosmological 
constant. The value of the ten-form at spatial infinity is the
charge of the domain wall at the corresponding asymptotic region.
 The main property of the D-8-brane
solution is that the cosmological constant changes 
discontinuously at the position of the brane.
We expect that domain walls on the worldvolume of branes will have similar
properties.  In particular, (i) there should be a field that changes 
discontinuously
at the position of the wall and (ii) the domain wall will separate the 
worldvolume
of the brane into two regions. In this work we shall investigate charged 
domain walls which are associated
with a (p+1)-form field strength on the worldvolume of the
p-brane. There are two kinds of worldvolume (p+1)-form field strengths: (i) 
one class of (p+1)-form field strengths is associated  
to the tension of branes in the
scale invariant formulation [\extra, \ericpaul] of the worldvolume brane 
actions\foot{ The substitution of a (p+1)-form field strength for the 
tension of a brane
is similar to the substitution of a ten-form field strength for the 
ten-dimensional cosmological
constant.}; (ii) 
another class of such worldvolume forms is due to intersections of a
 p-brane with a q-brane
on a (p-1)-brane. As it has been argued in [\pt] for such 
intersections to occur the
worldvolume action of the p-brane should 
have a (p+1)-form field strength in its spectrum. If the intersecting brane
configuration preserves a proportion of the supersymmetry of the bulk,
the associated worldvolume domain wall will preserve the same
proportion of supersymmetry.

In this paper, we first present domain wall solutions for all
 p-branes which are  associated with the first class of
(p+1)-form field strengths mentioned above, i.e. the ones that 
correspond to their 
tension in the scale invariant formulation
of the worldvolume actions. We shall show that 
these domain wall solutions  break 
all the supersymmetry of the bulk. We shall verify 
this both by a direct computation and
by an argument based on intersecting branes.
Next, we shall examine domain wall solutions on p-branes associated
with the second class  of
(p+1)-form field strengths.  We shall be mainly concerned
with the intersection of a IIB D-5-brane and a IIB NS-5-brane on a 4-brane.
>From either the perspective of the D-5-brane or the perspective of the
 NS-5-brane
the associated worldvolume soliton is a domain wall. 
We shall argue that the worldvolume actions of the D-5-brane 
and the NS-5-brane should contain a 6-form field strength due to the
intersection which belongs to the second class and is
different from the one that describes their tensions.
Using an analogy with  the standard BI field we shall give
the kappa-symmetry transformations of the actions involving this new
6-form field strength up to terms linear in the fields. 
We shall present a solution
for this domain wall and we shall show that it is supersymmetric.

This letter is organized as follows: In section two, we give
the domain wall solutions associated with the brane tensions and show
that they are not supersymmetric. In section three, we give two arguments
for the existence of supersymmetric domain walls; one based on a T-duality
chain and the other based on the six-dimensional
(1,1) worldvolume supersymmetry
algebra. In section four, we discuss such supersymmetric domain wall 
solutions for the
IIB NS-5-brane and IIB D-5-brane and in section five we present our 
conclusions.  
In a separate appendix, following [\ericpaulb], 
we list all the central charges of the supersymmetry 
algebra corresponding to the IIB NS-5-brane and relate every soliton on the 
worldvolume of the IIB NS-5-brane to a possible intersection involving a 
IIB NS-5-brane in the bulk.

\chapter{Domain Walls and Scale Invariant Actions}

The domain walls that we shall investigate do not involve the standard 
Born-Infeld fields.  This
allows us to treat all cases in a uniform way. 
Let $X$ be the embedding map of a p-brane
into ten-dimensional Minkowski spacetime.  
We introduce a Lagrange multiplier $V$ and write
the worldvolume action of the p-brane as [\extra, \ericpaul]
$$
I=\int \, d^{p+1}u\, \big({1\over2V} {\rm det} (g_{\mu\nu})-T^2 V\big)\
,
\eqn\actiona
$$
where $T$ is a parameter which is identified with the tension of the p-brane, 
$\{u^\mu;
\mu=0,\dots, p\}$ are the worldvolume coordinates and  
$$
g_{\mu\nu}=\partial_\mu X^M \partial_\nu X^N\eta_{MN}
\eqn\y$$
is the induced metric on the p-brane ($\eta_{MN}$ is he ten-dimensional
Minkowski spacetime metric). Using
the field equation of the Lagrange multiplier, it is straightforward
 to show that this action is
equivalent to the usual Nambu-Goto action. Next, 
we shall follow [\ericpaul] and postulate the
action
$$
I=\int \, d^{p+1}u \big({1\over2V} \big[{\rm det} (g_{\mu\nu})+4
(\tilde F)^2\big]\big)\ ,
\eqn\actionb
$$
where
$$
F_{\mu_1,\mu_2,\dots,\mu_{p+1}}=(p+1) 
\partial_{[\mu_1} A_{\mu_2,\dots,\mu_{p+1}]}
\eqn\z
$$
is a (p+1)-form field strength and 
$$
\tilde F={1\over (p+1)!} \epsilon^{\mu_1,\dots, \mu_{p+1}} 
F_{\mu_1,\dots,\mu_{p+1}}
\eqn\a
$$
is the Poincar\'e dual of $F$.

The field equations of the above action are
$$
\eqalign{
{\rm det} (g_{\mu\nu})+4
(\tilde F)^2&=0\, ,
\cr
\partial_\mu (V^{-1} \tilde F)&=0\, ,
\cr
\partial_\mu (V^{-1} {\rm det} (g_{\lambda\rho}) g^{\mu\nu} \partial_\nu 
X^M)&=0\, .}
\eqn\fielde
$$
It has been shown in [\ericpaul] that the field equations of the action 
\actiona\ are the same as
those  of the action \actionb. To see this, we solve the second equation 
in \fielde\ by setting
$$
V^{-1} \tilde F={T\over \sqrt 2} \ ,
\eqn\tension
$$
where $T$ is the tension of the p-brane as in \actiona.

To find the domain wall worldvolume soliton on a p-brane associated 
with $F$, we shall adopt
a similar point of view to that adopted for finding the D-8-brane 
solution of the
massive IIA supergravity theory [\eric] and allow the tension $T$ to be 
{\sl piece-wise constant}. 
Next, we choose the static gauge,
$\{X^M\}=\{ (u^\mu, Y^i); \mu=0,\dots,p,  i=p+1, \dots, 9\}$. 
We take $z=u^p$ to be the
coordinate transverse
to the domain wall and write the ansatz
$$
\eqalign{
Y^i&=Y^i(z)\, ,
\cr
V&=V(z)\, ,
\cr
\tilde F&=(-1)^p \partial_z W(z)\, .}
\eqn\b
$$
 Using this ansatz, the last two field equations
of \fielde\ can be rewritten as
$$
\eqalign{
\partial_z (V^{-1} \partial_z W)&=0\, ,
\cr
\partial_z (V^{-1} \partial_z Y^i)&=0\, .}
\eqn\fieldetwo
$$
One solution of these  equations is
$$
V^{-1} \partial_z W=\cases{w_1\, \ {\rm for}\, \  -\infty <z<0 \cr
                  		w_2\, \ {\rm for}\,\,\, 0<z<+\infty\ ,}
\eqn\solone
$$
and
$$
V^{-1} \partial_z Y^i=\cases{y^i_1\, \ {\rm for}\, -\infty <z<0 \cr
                  		y^i_2\, \ {\rm for}\,\,\, 0<z<+\infty\ ,}
\eqn\soltwo
$$
where $w_1, w_2, y^i_1, y^i_2$ are real constants. 
Substituting \solone\  and \soltwo\ into the first field equation
of \fielde\  we
find
$$
V(z)=\cases{\pm {1\over \sqrt {4 w_1^2- |y_1|^2}}\, \ 
{\rm for}\, -\infty <z<0 \cr
                  		\pm {1\over \sqrt {4 w_2^2- |y_2|^2}}\, \ 
{\rm for}\,\,\, 0<z<+\infty\ ,}
\eqn\c
$$
where $|y_1|^2=\delta_{ij} y^i_1 y^j_1$ and $|y_2|^2=\delta_{ij} y^i_2 y^j_2$.
It remains to solve for $W$ and $Y$. 
If we allow $V$ to be piece-wise constant, then the
equations \fieldetwo\ will become differential equations with 
distributional coefficients.
We wish to avoid this, so we shall take $V$ to be constant 
everywhere which leads to
the condition
$$
4 w_2^2- |y_2|^2=4 w_1^2- |y_1|^2
\eqn\con
$$
for the constants of the solution.  
The equations \fieldetwo\ then imply that $W, Y^i$ are harmonic
functions which we shall take to be continuous but not differential at 
$z=0$, {\sl i.e.}
$$
W(z)=\pm {1\over \sqrt {4 w_1^2- |y_1|^2}} \cases{w_1 z\, \ 
{\rm for}\, \ -\infty <z<0 \cr
                  		w_2 z\, \ 
{\rm for}\, \  0<z<+\infty\ ,}
\eqn\d
$$
and
$$
Y^i=\pm {1\over \sqrt {4 w_1^2- |y_1|^2}} \cases{y^i_1 z\, \
{\rm for}\,  -\infty <z<0 \cr
                  		y^i_2 z\, \ {\rm for}\, \  0<z<+\infty\ .}
\eqn\e
$$
A more general solution can be found by setting
$$
\eqalign{
Y^i&=H_2^i\, ,
\cr
\tilde F&=(-1)^p \partial_z H_1\, ,}
\eqn\h
$$
where $H_1, H_2^i$ are harmonic functions of the real line.
For $V$ to be a constant, as required, $H_1, H_2^i$ should have the same 
centres and their
first derivatives should satisfy \con\ at every centre. 
The domain walls are located
at the centres of the harmonic functions. Using the
relation between the tension and
$\tilde F$, see \tension, we find that
$$
T=(-1)^p {\sqrt 2} V^{-1} \partial_z H_1\ .
\eqn\i
$$
It is clear from this that the tension of the p-brane is different at 
the two sides of a domain wall.

These domain walls are not supersymmetric. This can be 
verified by a direct computation  using the
kappa-symmetry transformations of the scale invariant 
brane actions as given in [\ericpaul], {\sl i.e.} 
$$
\delta \theta= V^{-1}(\tilde F+ \Gamma_{(0)}) \kappa\ ,
\eqn\j
$$
where $\kappa$ is the transformation parameter and $\Gamma_{(0)}$ is the 
product structure
associated with the p-brane (see  e.g.
 [\papac] where it has been been denoted with
$\Gamma_{(0)}'$). The above kappa-symmetry transformation will suffice for 
our purposes since the
standard BI field vanishes for our configurations. Using a modification of the 
argument in [\papac], we
find that the supersymmetry condition is
$$
(\tilde F- \Gamma_{(0)})\epsilon=0\ ,
\eqn\susycon
$$
where $\epsilon$ is the supersymmetry parameter.
We shall not give the
details of this computation for all p-branes but we shall consider the case 
of the D-2-brane.
Let us suppose that the D-2-brane lies in the direction $0,1,2$ and 
that there is one non-vanishing 
transverse scalar along the direction $3$. Substituting the 
domain wall ansatz in
\susycon, we find that
$$
\big(\partial_z W- \Gamma_0\Gamma_1 (\Gamma_2+\partial_z Y \Gamma_3)\big)
\epsilon=0\ .
\eqn\k
$$
It is clear that for the solution to be supersymmetric one should impose 
the conditions
$$
\eqalign{
\Gamma_0\Gamma_1 \Gamma_2\epsilon&=\epsilon\, ,
\cr
\Gamma_0\Gamma_1 \Gamma_3\epsilon&=\epsilon\, .}
\eqn\prodcon
$$
However there are no non-trivial solutions to the above equations since
 the two product structures
anticommute and therefore do not have common solutions. 

 An alternative way to show that such a
domain wall soliton is not supersymmetric is to use the bulk picture of such 
solitons as arising
from
 the intersection of two branes.
We shall present our argument  for the case that only one of the 
transverse scalars $Y$ is non-zero. We remark that for these solutions 
$\tilde F$ does not vanish. Let us suppose that there is a supersymmetric 
intersecting brane
configuration associated with the above domain wall solitons. Since the 
soliton changes the tension
of the p-brane discontinuously, it implies that the intersection will involve
two p-branes  of the same kind intersecting on a (p-1)-brane.  However no 
such supersymmetric bulk
intersecting brane configurations exist which  preserve $1/4$ or less of 
the supersymmetry of the
bulk. This argument is consistent with the example that we have presented 
above. For this, observe
that the product structures of \prodcon\ are those of two planar D-2-branes 
lying in the directions
$0,1,2$ and $0,1,3$, respectively, and therefore 
intersecting at a string. There is no such
supersymmetric configuration.

\chapter{Domain Walls and Supersymmetry}

So far we have investigated non-supersymmetric domain walls. Here we shall 
present
two arguments indicating that there exist  supersymmetric domain walls 
as well. One of the arguments is based
on T-duality and the other is based on the properties
of the worldvolume supersymmetry algebra.

To give some examples of  supersymmetric domain walls, we first
extend the T-duality chain of 
[\papanew]  as follows:
$$
\eqalign{
(0|1_D,5_S)_B&{\buildrel  T\over\leftrightarrow}
(1|2_D,5_S)_A {\buildrel  T\over\leftrightarrow}
(2|3_D,5_S)_B{\buildrel  T\over\leftrightarrow}
\cr
(3|4_D,5_S)_A&{\buildrel  T\over\leftrightarrow}
(4|5_D,5_S)_B{\buildrel  T\over\leftrightarrow}
(5|6_D,5_S)_A\ .}
\eqn\chainone
$$
The notation that we are using, $(k|p,q)$, denotes a p-brane and a q-brane 
intersecting on
a k-brane and the various subscripts are self-explanatory.
As suggested in [\papanew], from the perspective of NS-5-branes the various 
brane solitons lying on
the  intersections of two branes or 
as boundaries of branes ending on branes 
transform like D-branes under IIA and IIB T-duality.
For the first four intersections of the T-duality chain a corresponding
supersymmetric worldvolume soliton has been constructed (for the
IIB 0-brane soliton, 
see [\gibbons]; for the IIA string and three-brane solitons, 
see [\howeb, \howe]). 
The next intersection
of the T-duality chain naturally suggests the existence of a 
4-brane, i.e. domain wall solution, on the NS-5-brane. 
Since T-duality preserves supersymmetry, we expect the domain wall solution
to be supersymmetric. 
This domain wall
solution will be analyzed in the next section. 

Let us consider the case $(4|5_D,5_S)_B$ in more detail. 
 From either the perspective of the IIB D-5-brane or the
perspective of the IIB NS-5-brane the 4-brane soliton is a worldvolume 
domain wall.
>From this D-brane perspective,
the above duality chain suggests that there should be
a {\sl supersymmetric} domain wall solution on every D-brane.

An alternative way to argue for the existence of supersymmetric domain walls
on the IIB NS-5-brane and IIB D-5-brane is to use their worldvolume 
supersymmetry
algebra as in [\hull, \ericpaulb].
The worldvolume theory of both IIB NS-5- and D-5-branes is described by the 
reduction to six
dimensions of the N=1 ten-dimensional super-Maxwell multiplet. The associated 
six-dimensional (1,1)
supersymmetry algebra including central charges is given by
$$
\eqalign{
\{Q_{\alpha i}, Q_{\beta j}\}&=\epsilon_{ij} (\gamma^\mu)_{\alpha\beta} 
(P+Z)_\mu +
(X^+)_{(\alpha\beta) (ij)}
\cr
\{Q_{\alpha i}, \tilde Q^\beta{}_{\underline j}\}&=
(\gamma^{\mu\nu})_\alpha{}^\beta Y_{\mu\nu,
i{\underline j}}+ W_{i{\underline j}}
\delta_\alpha{}^\beta
\cr
\{\tilde Q^\alpha{}_{\underline i}, \tilde Q^\beta{}_{\underline j}\}&=
\epsilon_{{\underline
i}{\underline j}} (\gamma^\mu)^{\alpha\beta} (P-Z)_\mu +
(X^-)^{(\alpha\beta)}{}_{({\underline i}{\underline j})}\ ,}
\eqn\susyalg
$$
where $\{Q_{\alpha i}, \tilde Q^\alpha{}_{{\underline i}}; \alpha=1,\dots, 4;
 i, {\underline i}=1,
\dots, 2\}$ are the sixteen supersymmetry charges, $P$ is the energy momentum, 
$Y_{ i{\underline j}}$ are two-form charges, $Z$ is a one-form  charge, 
$W_{i{\underline j}}$ are
scalar charges and
$X^+, X^-$ are the self-dual and the anti-self-dual parts of a three-form 
charge\foot{The greek
indices are spinor indices while the Roman ones transform under the
automorphism group  
$SU(2)\times SU(2)$
 of the algebra which is the spin group of $SO(4)$.}.
Following [\ericpaulb], we identify the central charges of the supersymmetry 
algebra with the
various brane solitons on the worldvolume of the associated branes
(for a full list and the relation to all bulk intersections involving the IIB 
NS-5-brane, see the appendix). 
In particular, from the
perspective of the NS-5-brane the charge that corresponds to the domain wall 
soliton is the
`electric' component of the two-form central charge
$Y_{i{\underline j}}$ which is a vector of $SO(4)$.  It is straightforward 
to show that
this soliton is supersymmetric using the supersymmetry algebra.

\chapter{Supersymmetric Domain Walls on IIB 5-branes}

In this section we shall investigate the supersymmetric domain wall 
solutions on the IIB 
NS-5-brane. The
related problem of domain wall solutions on the IIB D-5-brane is S-dual to 
this.
In order to have domain wall solutions on the IIB NS-5-brane, 
the worldvolume action of the 
theory
should include a 6-form field strength $f$\foot{Since we are interested
in supersymmetric domain walls we assume that this is a 
different 6-form field strength 
from that of the scale invariant action considered in section two.}. 
However such a  worldvolume
action for the NS-5-brane is not known\foot{This is related to the problem
of how to construct $SL(2,\R)$ covariant actions for IIB (p,q) 5-branes.}. 
The presence of the 6-form field strength $f$ on the 
worldvolume of
the NS-5-brane is due to the intersection of the IIB NS-5-brane with the IIB 
D-5-brane. This is
reminiscent to the presence of the standard BI field in the worldvolume 
action of  D-branes 
because fundamental strings end on D-branes\foot{For the case
of IIB (p,q) 1-branes a
$SL(2,\R)$ covariant action, containing {\sl two} worldvolume vector fields,
has been constructed [\paulmartin].}.
Using this analogy, we write an action for the IIB NS-5-brane up to quadratic 
terms for the fields
which includes
$f$ as
$$
I={T\over 2} \int d^6u \big( \eta^{\mu\nu} \partial_\mu Y^i 
\partial_\nu Y^j + |f|^2\big)
\eqn\l
$$
in the static gauge. 
Following the analogy, the kappa-symmetry transformations of this new action
should be similar to those of D-branes including the contribution from the 
(standard)
BI field. So we write the kappa-symmetry transformation of the spacetime 
fermions $\theta$ as\foot{Note that this kappa-symmetry rule differs
from the one used in section 2.}
$$
\delta \theta= [1+(1+  i \tilde f \sigma_2) \Gamma_{(0)}] \kappa\ ,
\eqn\m
$$
where $\kappa$ is the parameter, $\Gamma_{(0)}$ is the supersymmetry
projector associated with the IIB NS-5-brane and $\tilde f$ is the dual of 
$f$. In the above
expression, we keep only terms linear in the fields. The presence of 
$\sigma_2$ in the
kappa-symmetry transformations interchanges the two chiral ten-dimensional 
fermions
$\kappa$. To explain this, we use the
fact that
$f$ is associated with the intersection of the D-5-brane with the NS-5-brane
 and that the
supersymmetry projector of the D-5-brane interchanges the two chiral 
ten-dimensional fermions
$\kappa$ which serve as the parameters in the kappa-symmetry transformations 
of the D-5-brane
[\ericpaulc-\aga]. The associated supersymmetry condition is
$$
(1+  i \tilde f \sigma_2) \Gamma_{(0)}\epsilon=\epsilon
\eqn\n
$$
where $\epsilon$ is the supersymmetry parameter.

Next let us suppose that the IIB NS-5-brane is located in the directions
 $0,1,2,3,4,5$ and the
direction transverse to the domain wall is $z=u^5$. To
find a supersymmetric domain wall solution, we use the ansatz
$$
\eqalign{
\tilde f&=-\partial_z W\, , 
\cr
Y&=Y^6(z)\, .}
\eqn\o
$$
We remark that from the bulk perspective the D-5-brane lies in the directions
 $0,1,2,3,4,6$.
Substituting this ansatz in the supersymmetric condition above and keeping
 only terms
linear in the fields, we find that supersymmetry is preserved if
$$
\partial_z W=\partial_zY
\eqn\p
$$
provided that
$$
\eqalign{
\Gamma_0\Gamma_1\Gamma_2\Gamma_3\Gamma_4\Gamma_5\epsilon^1&=\epsilon^1\, ,
\cr
\Gamma_0\Gamma_1\Gamma_2\Gamma_3\Gamma_4\Gamma_5\epsilon^2&=\epsilon^2\, ,
\cr
\Gamma_0\Gamma_1\Gamma_2\Gamma_3\Gamma_4\Gamma_6\epsilon^1&=\epsilon^2\, .}
\eqn\q
$$
These are the bulk supersymmetry projectors of the IIB NS-5-brane and the 
IIB D-5-brane. In
particular, the first two projectors are associated with the NS-5-brane and 
the latter one is
associated with the  D-5-brane. The supersymmetry preserved is $1/4$ of the 
bulk and so $1/2$
of the worldvolume.
The field equations imply that $W$ is a harmonic function $H$ of $z$ and 
therefore the solution
can be written as
$$
\eqalign{
\tilde f=-\partial_z H 
\cr
Y=H(z)\ .}
\eqn\r
$$
It is straightforward now to find the supersymmetric domain wall on the IIB 
D-5-brane
due to the intersection with the NS-5-brane. 

It remains to comment about the possibility of domain walls on the other 
D-branes
involved in the T-duality chain  \chainone. We remark that in some cases, 
like the
one involving the D-string and IIB NS-5-brane, the D-brane ends on the 
NS-5-brane.
In such a case, it is not clear to us that there should be a domain wall 
from the
perspective of the D-brane\foot{A three-brane domain wall on the D-4-brane 
worldvolume has been discussed in [\howe].}. 

\chapter{Conclusions}

The worldvolume actions of p-branes admit domain wall solutions which are
associated with various worldvolume
(p+1)-form field strengths. First, we have investigated
a class of non-supersymmetric domain walls where the worldvolume
(p+1)-form field strength
is dual to the tension of the p-brane. Next, we we have given one example of 
supersymmetric
domain walls on the IIB NS-5- and D-5-branes. The T-duality argument
presented in section three  suggests the existence of 
many more supersymmetric
domain wall solutions which we plan to investigate elsewhere.

One application of the existence of domain walls on the
worldvolume of branes is to indicate that there are processes where the brane
is separated into two pieces along the wall. Such processes will
involve the tension of the brane and therefore the non-supersymmetric
domain walls.

Another implication of our work is that the worldvolume
(p+1)-form field strength, needed in the construction of the domain wall
solution, suggests that the maximal supersymmetric Maxwell 
multiplet
in  D-dimensions admits an extension by a D-form field strength. Such a field
does not introduce additional propagating degrees of freedom and similar 
extensions
have been considered before
 for other multiplets [\gates]. This follows from 
the fact that all D-branes have a kappa-symmetric scale invariant 
action\foot{We thank Paul Townsend for a discussion on this point.}
.
We shall demonstrate this for the ten-dimensional Maxwell theory. 
The other
cases follow by dimensional reduction. Let $A$ be the one-form Maxwell gauge
 potential, $F=dA$,
$\chi$ be the spacetime Majorana-Weyl fermion partner of $A$ and $B$ be a 
9-form gauge potential,
$G=dB$.  An action for this multiplet is
$$
S = \int d^{10}x\ |F|^2+\bar \chi \dpar\chi+ |\tilde G|^2\ ,
\eqn\s
$$
where $\tilde G$ is the dual of $G$.
The supersymmetry transformations, leaving this action invariant, are given by 
$$
\eqalign{
\delta A&=\bar \epsilon\dpar\chi 
\cr
\delta \chi&= \Gamma^{\mu\nu} F_{\mu\nu}\epsilon+ \tilde G \eta
\cr
\delta B_{\mu_1, \dots, \mu_9}&=\bar\eta \Gamma_{\mu_1, \dots, \mu_9} \chi\ .}
\eqn\t
$$
where $\epsilon$ is the standard supersymmetry parameter and $\eta$ is a new 
parameter that has the same chirality as $\epsilon$. Note that the 
supersymmetry algebra closes up to a central charge transformation.
It would be interesting to investigate the properties of this modified
Maxwell multiplet further.

 \vskip 1cm

\noindent{\bf Acknowledgments:}  We thank G.W. Gibbons for helpful
discussions. G.P. thanks the Institute for 
Theoretical Physics of
the University of Groningen for hospitality.  G.P. is supported by a 
University Research Fellowship from the Royal Society.
This work is part of the Research program of the
``Stichting voor Fundamenteel Onderzoek der Materie''(FOM).
It is also supported by the European Commission TMR programme
ERBFMRX-CT96-0045, in which E.B. is associated
to the University of Utrecht.

\vfill\eject

\appendix

In this appendix, following [\ericpaulb], 
we list all the central charges of the supersymmetry 
algebra corresponding to the IIB NS-5-brane and relate every soliton on the 
worldvolume of the IIB NS-5-brane to a possible intersection involving a 
IIB NS-5-brane in the bulk.  

As we have already mentioned in section three, the supersymmetry algebra 
\susyalg\
has two three-form charges $X^+, X^-$, a two-form charge 
$Y^{i{\underline j}}$,  two one-form
charges
$P, Z$ and a scalar charge $W_{i{\underline j}}$. These charges transform 
under $SO(4)$ as
follows: $X^+, X^-$ transform as self-dual, anti-self-dual two forms, 
respectively,
$Y^{i{\underline j}}, W_{i{\underline j}}$ transform as a vector and 
$Z$ transforms as a 
scalar.   We summarize in the following table the worldvolume solitons 
associated
with these central
charges as well as the number of non-zero transverse scalars associated
with them. We do this for the NS-5-brane. The table for the D-5-brane is 
similar.
\bigskip

$$
\vbox{\settabs 5\columns \+ {\rm Charge}& p & $p^\prime$& t&$t^\prime$\cr
\+&&&&\cr
\+ $X^+$&3&-&2&-\cr
\+ $X^-$&-&3&-&2\cr
\+ Y&2&4&1&3\cr
\+ Z&1&5&0&4\cr
\+ W&0&-&1&3\cr}
$$
\bigskip

\noindent {\bf Table 1: Worldvolume solitons.}\ \  This table relates
every central charge of the supersymmetry algebra \susyalg\ to a p-form 
( $p^\prime$-form) worldvolume soliton.
The numbers t and $t^\prime$ indicate the number of transverse directions
from the bulk perspective.
\bigskip

The brane intersections associated with these worldvolume solitons are 
summarized in
the following table:
\bigskip

$$
\vbox{\settabs 5\columns \+ {\rm Soliton} & t & $t^\prime$\cr
\+&&\cr
\+ 0& $(0|5_S,1_D)$& $(0|5_S,3_D)$\cr
\+ 1& $(1|5_S,1_F)$& $(1|5_S,5_S)$\cr
\+ 2& $(2|5_S,3_D)$& $(2|5_S,5_D)$\cr
\+ 3& $(3|5_S,5_S)$& $(3|5_S,KK)$\cr
\+ 4& $(4|5_S,5_D)$& $(4|5_S,7_D)$\cr
\+ 5& $(5|5_S,KK) $& $(5|5_S,9_S)$\cr}
$$
\bigskip

\noindent
{\bf Table 2: Brane intersections.}\ \  This table relates every p-form
($p^\prime$-form) worldvolume soliton of the previous table to two 
different brane intersections. The two cases correspond to the cases
t and $t^\prime$ in the previous table. 

\noindent
The intersections involving the D-5-brane can be summarized in a similar way.
\vskip 1cm

\refout

\bye